# *Circuit elements at optical frequencies: nano-inductors, nano-capacitors and nano-resistors*

*Nader Engheta*[(1),*], *Alessandro Salandrino*[(1,2)], *and Andrea Alù*[(1,2)]

(1) University of Pennsylvania, Department of Electrical and Systems Engineering, Philadelphia, PA, U.S.A., engheta@ee.upenn.edu

(2) University of Roma Tre, Department of Applied Electronics, Rome, Italy, alu@uniroma3.it

**Abstract**

We present some ideas for synthesizing nanocircuit elements in the optical domain using plasmonic and non-plasmonic nanoparticles. Three basic circuit elements, i.e., nano-inductors, nano-capacitors, and nano-resistors, are discussed in terms of small nanostructures with different material properties. Coupled nanocircuits and parallel and series combinations are also envisioned, which may provide road maps for the synthesis of more complex nanocircuits in the IR and visible bands. Ideas for the optical implementation of right-handed and left-handed nano-transmission lines are also forecasted.

---

[*] To whom correspondence should be addressed

Optical wave interaction with metallic and non-metallic nanoparticles is one of the important problems in nanotechnology and nano-photonics nowadays. It is well known that in certain noble metals such as Ag, Au, the plasma frequency is in the visible or ultraviolet (UV) regimes, and thus these metals behave as plasmonic materials in the optical frequencies, i.e., their permittivity has a negative real part [1]. As a result, the interaction of optical signals with plasmonic nanoparticles involves surface plasmon resonances [e.g., 1,2]. Since these particles may be much smaller than the optical wavelength, the following question may naturally arise: 'Can such metallic (and non-metallic) nanoparticles be treated as nanocircuit "lumped elements", such as nano-inductors, nano-capacitors, and nano-resistors in the optical regimes?'. The conventional circuits in the lower frequency domains (such as in the RF and lower frequency range) indeed involve elements that are much smaller than the wavelength of operation, and the circuit theory may be regarded as the "approximation" to the Maxwell equations in the limit of such small size. Similarly, in the following we explore quantitatively how these circuit concepts and elements may be extended to the optical frequencies when dealing with nanoparticles. It should be pointed out that a mere scaling of the circuit component concepts conventionally used at lower frequencies may not work at frequencies beyond the far infrared, since conducting materials (e.g., metals) behave differently at these higher frequencies. Here, however, we present a concept to synthesize circuit elements, and more complex circuits, in the optical regime.

To begin, we consider a nanosphere of radius $R$ made of a homogeneous material with dielectric function $\varepsilon(\omega)$, which is in general a complex quantity. The sphere is assumed to be much smaller than the wavelength of operation in vacuum, i.e., $R \ll \lambda_0$. Consider an incident electromagnetic wave $\mathbf{E}_0$ illuminating this sphere under a monochromatic excitation $e^{-i\omega t}$. Due to the small size of the particle (w.r.t. wavelength), the scattered electromagnetic fields in the vicinity of the sphere and the total fields inside it may be obtained with very good approximation using the well known time-harmonic, quasi-static approach. This leads to the following approximate expressions for the fields inside and outside the sphere [3]:

$$\mathbf{E}_{int} = \frac{3\varepsilon_0}{\varepsilon + 2\varepsilon_0}\mathbf{E}_0, \qquad (1)$$

$$\mathbf{E}_{ext} = \mathbf{E}_0 + \mathbf{E}_{dip} = \mathbf{E}_0 + \frac{3\mathbf{u}(\mathbf{p}\cdot\mathbf{u}) - \mathbf{p}}{4\pi\varepsilon_0 r^3}, \qquad (2)$$

with $\mathbf{p} = 4\pi\varepsilon_0 R^3 \frac{\varepsilon - \varepsilon_0}{\varepsilon + 2\varepsilon_0}\mathbf{E}_0$, $\mathbf{u} = \mathbf{r}/r$, $\varepsilon_0$ being the permittivity of the outside region, $\mathbf{r}$ being the position vector from the sphere's center to the observation point, and $r = |\mathbf{r}|$. At every point on the surface of the sphere the normal component of the displacement current $-i\omega D_n$ is continuous, implying that:

$$-i\omega(\varepsilon - \varepsilon_0)\mathbf{E}_0 \cdot \hat{\mathbf{n}} = -i\omega\varepsilon_0 \mathbf{E}_{dip} \cdot \hat{\mathbf{n}} + i\omega\varepsilon \mathbf{E}_{res} \cdot \hat{\mathbf{n}}, \qquad (3)$$

where $\hat{\mathbf{n}}$ is the local outward unit vector normal to the surface of the sphere. In this equation, $\mathbf{E}_{res} \equiv \mathbf{E}_{int} - \mathbf{E}_0$ represents the residual field internal to the nanosphere when the incident field is subtracted from the total internal field.

If $\mathbf{E}_0$ is oriented as shown by the black arrows in Fig. 1, when Eq. (3) is integrated over the upper hemispherical surface we get the "total" displacement current for each relevant term in Eq. (3):

$$\underbrace{-i\omega(\varepsilon-\varepsilon_0)\pi R^2|\mathbf{E}_0|}_{I_{imp}} = \underbrace{-i\omega\varepsilon\pi R^2\frac{\varepsilon-\varepsilon_0}{\varepsilon+2\varepsilon_0}|\mathbf{E}_0|}_{I_{sph}}\underbrace{-i\omega\varepsilon_0 2\pi R^2\frac{\varepsilon-\varepsilon_0}{\varepsilon+2\varepsilon_0}|\mathbf{E}_0|}_{I_{fringe}}. \qquad (4)$$

We have named the three terms in Eq. (4) according to their function: the "impressed displacement current source" $I_{imp}$, the "displacement current circulating in the nanosphere" $I_{sph}$, and the "displacement current of the fringe (dipolar) field" $I_{fringe}$, respectively. All of them are related to the polarization charges on the surface of the nanosphere, induced by the excitation. The above relationship among the various segments of the displacement current can be interpreted as the branch currents at a node in a parallel circuit, as shown in Fig. 1. Indeed, such currents as defined above obey the Kirchhoff current law, represented by Eq. (4). The Kirchhoff voltage law is also satisfied, since $\nabla \times E$ is locally zero in this quasi-static approximation.

The equivalent impedance for the "nanosphere" and the "fringe" branches of the circuit shown in Fig.1, are calculated as the ratio between the "average" potential difference (due to $\mathbf{E}_{res}$) between the upper and lower hemispherical surfaces of the sphere

$$\langle V\rangle_{sph} = \langle V\rangle_{fringe} = R\frac{\varepsilon-\varepsilon_0}{\varepsilon+2\varepsilon_0}|\mathbf{E}_0| \qquad (5)$$

and the effective currents evaluated in Eq. (4). Thus, we get:

$$Z_{sph} = (-i\omega\varepsilon\pi R)^{-1}, \quad Z_{fringe} = (-i\omega 2\pi R\varepsilon_0)^{-1}. \qquad (6)$$

From Eq. (6) we can clearly see that the two parallel elements in the circuits shown in Fig. 1 may behave differently according to the sign of the nanosphere's permittivity. Let us consider the following two cases:

*Non-metallic (i.e., non-plasmonic) sphere as a nano-capacitor*: In this case, the real part of $\varepsilon$ is a positive quantity, and thus $Z_{sph}$ in Eq. (6) is capacitive along with the resistive part related to the imaginary part of permittivity. The impedance of the outside fringe is always capacitive, since we assume that the permittivity of the outside region is positive. Thus the equivalent nano-circuit for a non-plasmonic nanosphere, which is small compared with the optical wavelength, can be shown as in the bottom left part of Fig. 1. Here the equivalent circuit elements can be expressed in terms of parameters of the nanospheres as follows:

$$C_{sph} = \pi R \operatorname{Re}[\varepsilon], \qquad G_{sph} = \pi \omega R \operatorname{Im}[\varepsilon], \qquad C_{fringe} = 2\pi R \varepsilon_0 \qquad (7)$$

Since there are two capacitive elements, there is no resonance present in this case – a fact that is consistent with the absence of resonance for optical wave interaction with the small non-plasmonic nanosphere.

*Metallic (i.e., Plasmonic) sphere as a nano-inductor*: In this case, we assume the sphere to be made of a plasmonic material, such as noble metals in the visible or IR band (e.g., Ag, Au), and as a result the real part of the permittivity may attain a negative value in these frequency bands. Therefore, the equivalent impedance of the nanosphere (Eq. 6) can be "negatively capacitive", which implies that at any given frequency for which $\operatorname{Re}[\varepsilon] < 0$, the equivalent capacitance is "negative". This can be interpreted as a positive *effective* "inductance", as discussed in [4-6]. Therefore, the equivalent circuit for the case of optical wave interaction with a plasmonic nanosphere may be presented as in the bottom right of Fig. 1. Here the equivalent circuit element for the sphere becomes:

$$L_{sph} = \left(-\omega^2 \pi R \operatorname{Re}[\varepsilon]\right)^{-1}. \qquad (8)$$

In this case, since there is an inductor in parallel to the fringe capacitor, the circuit may exhibit resonance, which corresponds to the plasmonic resonance for the optical wave interaction with the metallic nanoparticles, as mentioned in [4]. It may be verified that the resonant condition for the circuit $L_{sph}C_{sph} = \omega^{-2}$ requires the well known condition of plasmonic resonance for a nanosphere $\text{Re}[\varepsilon] = -2\varepsilon_0$ [1].

It follows from the above discussion that a small nanosphere excited by an optical signal may effectively behave as a "nano-capacitor" or a "nano-inductor" at the optical frequency, if the sphere is made of non-plasmonic or plasmonic materials, respectively. The imaginary part of the material permittivity may provide an equivalent nano-resistor. It is interesting to note that unlike the conventional design for an inductor in the lower frequency regimes, where the inductor is usually in the form of "wound wires", here this nano-element is made of a simple geometry consisting of plasmonic materials. In other words, instead of winding wires with dimensions much smaller than the optical wavelength, here the plasmonic characteristics of natural noble metals provide us with an effective inductance, whose value can be designed by properly selecting the size, shape, and material contents of the nanostructure. This indeed provides new possibilities for miniaturization of electronic circuits operating at the optical frequencies. The conventional circuits in the RF and lower frequencies, relying on the conduction current circulating in metallic wires along the lumped elements, cannot be straightforwardly scaled down to the infrared and optical frequencies, at which conducting metallic materials behave quite differently. However, introducing plasmonic and non-plasmonic nanoparticles as basic elements of optical nanocircuits, in which effectively the "*displacement*" current can similarly "circulate", may provide analogous functionalities at the optical frequencies. One may essentially have the three basic circuit elements, i.e., nano-inductor, nano-capacitor, and nano-resistor, operating in the optical frequency, which form the building blocks for the design of more complex nanocircuits at these wavelengths. To get an idea about the values of these nanoelements, let us assume a nanosphere with $R = 30\,nm$ made of silver. At the wavelength $\lambda_0 = 633\,nm$, the permittivity of silver is known to be $\varepsilon_{Ag} = (-19 + i0.53)\varepsilon_0$ [7]. From Eq. (8), we can then find $L_{sph} \simeq 7.12\,femtoH$, $G_{sph} = 1.32\,mS$, and $C_{fringe} \simeq 1.67\,attoF$. If the sphere is made of Au$_2$S with permittivity $\varepsilon_{Au_2S} = 5.44\varepsilon_0$ at $\lambda_0 = 633\,nm$, the nanocapacitance of the sphere will be $C_{sph} \simeq 4.53\,attoF$.

We point out that for a given wavelength and a specific material the values of these nanoelements depend directly on the radius of the nanosphere. However, if one wants to have more flexibility (i.e., more degrees of freedom) in their design, one can use nanoparticles with different geometries, e.g., ellipsoidal nanoparticles where there are three geometrical parameters, corresponding to the three axes.

Expanding this concept to configurations with more than one nanoparticle, e.g., the case of two nanospheres with radii $R_1$ and $R_2$, permittivities $\varepsilon_1$ and $\varepsilon_2$, and with a certain distance $d$ apart, it can be shown that these configurations may be effectively treated as "coupled" nano-circuits, each representing one of the spheres (See Fig. 2). Each circuit in the figure includes the capacitive or inductive impedance of the given nanosphere, the capacitive impedance related to the fringe field, and the current source representing the impressed field on this sphere. However, in addition, each circuit also needs to have a "dependent" current source representing the influence of the field of other particle(s) on this sphere. In other words, the interaction among the particles here can be exhibited using such dependent sources. The value of each dependent current source in Fig. 2 may be explicitly derived in terms of the potential difference across the other nanosphere, in analogy with the previous formulas.

In order to form parallel or series circuit elements with these nanoparticles, one would need to juxtapose two (or more) of them very closely with specific orientations with respect to the illuminating electric field. Fig. 3 (top row) shows the geometry of a structure that consists of two tightly paired semi-cylinders of differing permittivities. (Here, for the sake of mathematical simplicity, lossless cylinders are considered.) The potential distribution around this fused structure, when illuminated with an electric field, provides useful information about its behavior as combined circuit elements. The middle panels present the potential distribution and equipotential surfaces for the two cases of electric field being parallel (left column) and perpendicular (right column) to the plane interface between the two halves with permittivities $\varepsilon$ and $-\varepsilon$. (Since the diameter of the semi-cylinders is assumed to be much smaller than the operating wavelength, an approximate time-harmonic quasi-static analysis is used for evaluating the potential distribution here. The detail of this analysis and some of the salient features of potential distributions, also in connection with this circuit analogy, will be extensively reported in a future publication. This sample result is only shown here.) We note that the equipotential surfaces near the fused cylinder in the left column of Fig. 3 become perpendicular to its outer surface, implying that the normal component of the total electric field is zero at this surface. However, there is indeed a certain potential difference between the top and bottom parts of the cylinder's surface. It is possible to show that as seen from the outside this fused structure might be regarded as a *parallel* resonant L-C circuit (which in fact has an infinite impedance at its resonance, and hence zero net current flowing into it), in parallel with the fringe capacitor, as depicted in the equivalent circuit in the bottom left column of Fig. 3. (Since the materials are assumed to be lossless, no equivalent resistor is present here, but it may be easily added.) In an analogous way, the fused semi-cylinders in the right column of this figure, having the external electric field perpendicular to the boundary interface between the two halves, may be regarded as a *series* resonant L-C circuit, as observed from the outside. This equivalent circuit is reported in the bottom right column of Fig. 3. In fact, as seen in the middle right panel, the equipotential surfaces in this case become parallel with the fused cylinder's surface, implying that the potential difference at the surface of this structure is effectively zero, whereas the displacement current flows in and out of it. The resonant behavior of these examples is present due to the particular choice of oppositely signed (but equal magnitude) permittivities for the two halves. However, different pairs would behave as non-resonant series or parallel elements, depending upon their pairing and orientation with the external excitation. Moreover, other geometries for nanostructures similarly paired may lead to analogous parallel and series configurations. For instance, Fig. 4 (top left) shows a nanostructure conceptually formed by rectangular blocks of plasmonic and non-plasmonic materials. When this structure is excited by a local electric field of an optical signal (e.g., by a near-field scanning optical microscope (NSOM)) the plasmonic and non-plasmonic "blocks" may act as nano-inductors and nano-capacitors (along with some nano-resistor), respectively, and the structure may thus operate as the more complex circuit depicted in the bottom left panel of Fig. 4. Such nanocircuits can indeed behave as plasmonic nano-barcodes and plasmonic data storage systems. Such a nanocircuit may also be configured as a "closed" loop, as shown in the right part of Fig. 4. When this optical closed circuit is excited by an NSOM at one point, we speculate that the displacement current along this loop may behave as the current in a circuit formed by equivalent inductors and capacitors. Since the nanotechnology and fabrication techniques for making nanostructures with different metallic and oxide segments are actively being investigated by various groups (see e.g., [8]), construction of our proposed nanocircuits is within the realm of possibility. Such optical nanocircuits can also be interfaced with biological elements, such as cells, molecules, when these elements can substitute one of the plasmonic or non-plasmonic elements in the circuit. We are currently exploring some of these concepts.

Finally, it is interesting to point out that since one can have such nano-inductors and nano-capacitors in the optical frequency, by properly arranging these elements one may form optical nano-transmission lines. If the arrangement involves series nano-inductors and shunt nano-capacitors, this will provide conventional (also known as right-handed (RH)) transmission lines in the optical frequency. However, if the shunt nano-inductors and series nano-capacitors are used, we may synthesize negative-index (or left-handed (LH)) transmission lines in the optical domain, similarly to what recently proposed in the microwaves [5,9]. This may lead to interesting sub-wavelength focusing effects in the optical frequencies, together with a road map for development of negative-refractive-index (or left-handed) materials in the IR and visible regimes. Fig. 5 shows the sketches of such RH and LH optical transmission lines. We are currently studying these problems.

This work was supported in parts by the U.S. Defense Advanced Research Projects Agency (DARPA) Grant number HR0011-04-P-0042 and by the U.S. Air Force Office of Scientific Research (AFOSR) Grant number F49PRE-03-1-0438. Andrea Alù has been partially supported by the 2004 SUMMA Graduate Fellowship in Advanced Electromagnetics. The authors thank Dr. Natalia Bliznyuk for useful discussion during the early stage of this work. Portion of this work was presented at the 2004 USNC-URSI National Radio Science Meeting, June 20-26, 2004, Monterey, CA.

*Figures*

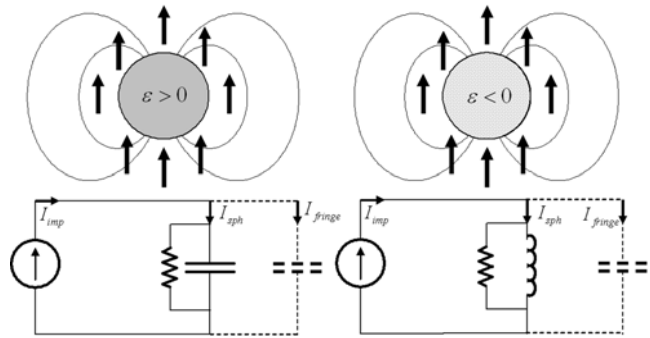

Fig. 1. A basic nanocircuit in the optical regime, using the interaction of an optical wave with an individual nanosphere. (left column) A non-plasmonic sphere with $\varepsilon > 0$, which provides a nano-capacitor and a nano-resistor; (right column) A plasmonic sphere with $\varepsilon < 0$, which gives a nano-inductor and a nano-resistor. Solid arrows show the incident electric field, and the thinner field lines represent the fringe dipolar field from the nanosphere.

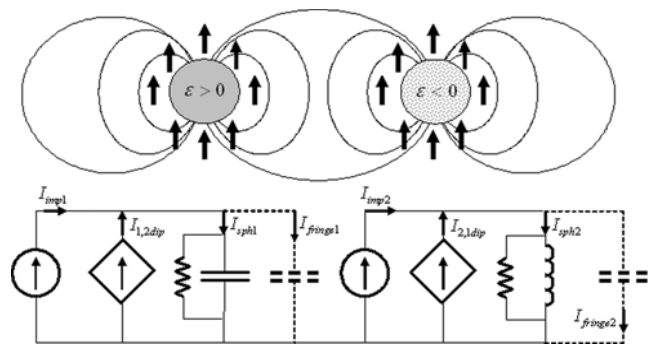

Fig. 2. A coupled nanocircuit in the optical domain, using optical wave interaction with two adjacent nanospheres.

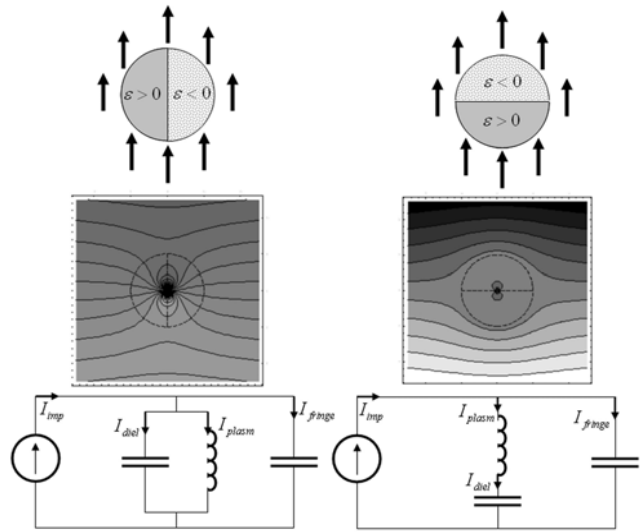

Fig. 3. Parallel and series nano-elements. (Top row): Two focused lossless semi-cylinders with positive and negative permittivities, illuminated by an optical field; (middle row): quasi-static potential distributions around and within the structure (solid lines show equipotential surfaces); (bottom row): Equivalent circuits showing parallel and series elements representing the fused structure as seen from the outside.

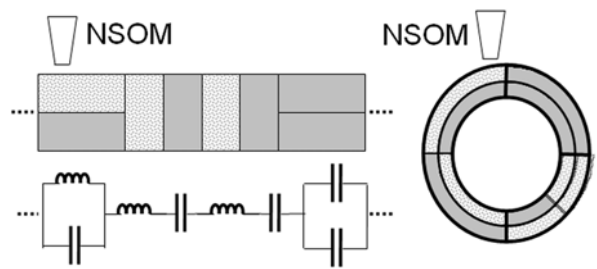

Fig. 4. Nanocircuit synthesis. (Top left) Conceptual nanocircuit formed by rectangular blocks of plasmonic and non-plasmonic segments; (bottom left) its equivalent circuit; (right) a closed "nano-loop".

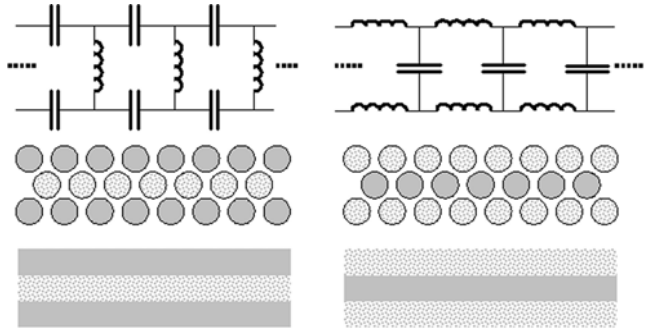

Fig. 5. An idea for optical implementation of right-handed (RH) and left-handed (LH) nano-transmission lines. (Top row): Conventional RH and LH lines using distributed (or lumped) inductor and capacitor elements; (middle row): Plasmonic and non-plasmonic nanostructures may play the role of nano-inductors and nano-capacitors; (bottom row): As the nanostructures gets closer, in the limit, plasmonic and non-plasmonic layers may be envisioned to constitute layered structures with forward and backward propagation properties.